\documentclass{vgtc}
\ifpdf
  \pdfoutput=1\relax                   
  \usepackage{graphicx}                
  \DeclareGraphicsExtensions{.pdf,.png,.jpg,.jpeg} 
\else
  \usepackage{graphicx}                
  \DeclareGraphicsExtensions{.eps}     
\fi%

\graphicspath{{figures/}{pictures/}{images/}{./}} 

\usepackage{microtype}                 
\PassOptionsToPackage{warn}{textcomp}  
\usepackage{textcomp}                  
\usepackage{mathptmx}                  
\usepackage{times}                     
\usepackage{cite}                      
\usepackage{tabu}                      
\usepackage{booktabs}                  

\usepackage{graphicx} 

\title{Refreshable Tactile Displays for Accessible Data Visualisation}
\author{Leona Holloway\\
    \scriptsize Monash University %
\and Peter Cracknell\\
    \scriptsize Quantum RLV
\and Kate Stephens\\
    \scriptsize Blind Consumer
\and Melissa Fanshawe\\
    \scriptsize University of Southern Queensland
\and Samuel Reinders\\
    \scriptsize Monash University
\and Kim Marriott\\
    \scriptsize Monash University
\and Matthew Butler\\
    \scriptsize Monash University}

\date{June 2023}
\abstract{Refreshable tactile displays (RTDs) are predicted to soon become a viable option for the provision of accessible graphics for people who are blind or have low vision (BLV). This new technology for the tactile display of braille and graphics, usually using raised pins, makes it easier to generate and access a large number of graphics. However, it differs from existing tactile graphics in terms of scale, height and fidelity. Here, we share the perspectives of four key stakeholders -- blind touch readers, vision specialist teachers, accessible format producers and assistive technology providers -- to explore the potential uses, advantages and needs relating to the introduction of RTDs. We also provide advice on what role the data visualisation community can take to help ensure that people who are BLV are best able to benefit from the introduction of affordable RTDs.}

\begin{document}

\CCScatlist{
\CCScatTwelve{Human-centered computing}{Accessibility}{Accessibility technologies}{}
\CCScatTwelve{Human-centered computing}{Visualisation}{}{}
}

\maketitle

\section{Introduction}
Tactile graphics, also known as raised line drawings, are considered best practice to convey spatial and graphical information for people who are blind or have low vision (BLV). Refreshable tactile displays (RTDs) are hardware devices with a moveable surface for the dynamic display of tactile graphics. They offer the opportunity for electronic storage, quick access to a large number of graphics, along with dynamic interaction and manipulation of content.

While RTDs are not new, they have remained a rarity due to their high price point. This will soon change, with several devices set to enter the market at significantly lower prices within the next two years. Those devices that look most promising for commercialisation, such as the DotPad~\cite{DotInc} and the Monarch~\cite{APH}, use refreshable pin displays with between 2,400 (60$\times$40) and 3,840 (96$\times$40) pins.
Additional features such as touch sensitivity and multiple pin heights are included in other devices such as the Graphiti~\cite{Orbit} and Metec~\cite{Metec}.

Here, we examine the stakeholder perspectives of how RTDs could be used to improve access to graphics, including those of data, by way of meaningful tactile data visualisations.

\begin{figure}[ht]
   \centering
 \includegraphics[width=0.5\textwidth]{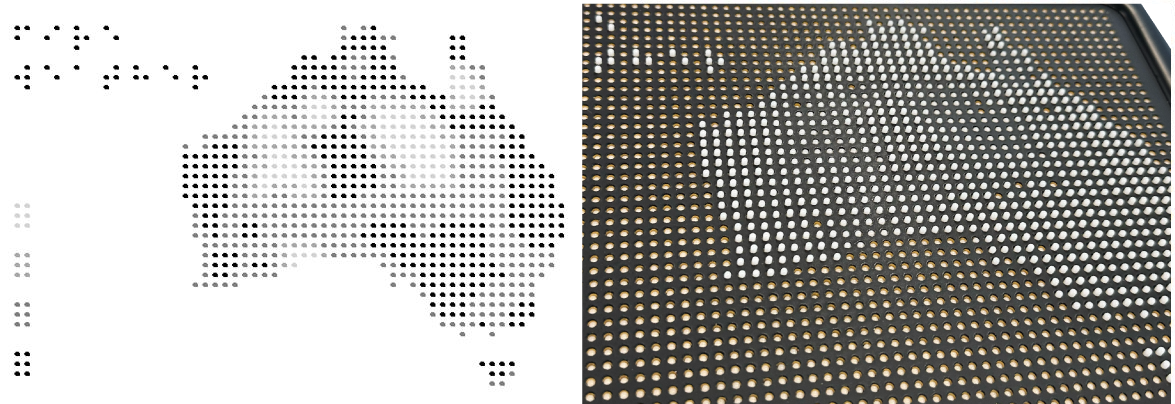}
 \caption{Pixelated heat map for display on an RTD (left) and the heat map displayed on a Graphiti RTD with variable pin heights (right)}
    \label{fig:heatmap}
\end{figure}

\section{Related Work}
Prior work into RTDs has mainly focused on haptics and perception~\cite{Garcia2011feeling}, zooming and panning~\cite{Ducasse2018botmap,Levesque2012adaptive},
-- with more work needed in this area -- and applications for using RTDs to display sports matches~\cite{Kobayashi2018football,Ohshima2021football} and maps~\cite{Brayda2018updated,Leo2020maps}.
There has also been a lot of interest in the potential use of RTDs for people who are BLV to create their own graphics as a tool for drawing~\cite{Bornshein2018drawing,Kobayashi2004pens} or for collaborative tactile graphics production~\cite{Bornshein2015collaborative}.

The display of data visualisations on RTDs has received much less attention. Visualisation of algebraic graphs has been the first application to be explored~\cite{Albert2006,Schwarz2022} but there are many more applications for extension within the data visualisation field.

\section {Stakeholder Perspectives}
The voices of four primary stakeholders presented here are based on our own experiences and opinions, combined with the perspectives of colleagues and collaborators within the print accessibility community. We are: a blind touch reader who has experience reading traditional tactile graphics and using an RTD; a vision specialist teacher who is also a parent of a student with a vision impairment; an accessible formats producer with experience in producing tactile graphics and contributing to guidelines for tactile graphics design; and an assistive technology provider with expertise in the range of assistive technology devices for touch access and their use.

We have retained the first person voice to emphasise the personal nature of our ideas but also in the hope that we can bring forward the voice of stakeholders as co-designers at the beginning of the design process for a new technology.

\subsection{Blind Touch Readers}
I am excited by the many possibilities that RTDs offer.

\emph{Dynamic content:} RTDs introduce the possibility of tactile animations~\cite{Holloway2022Animations}, especially if the refresh rate can be improved.

\emph{Timeliness:} The ability to have images updated in real-time and/or materials to be produced instantaneously.

\emph{User creation:} Most importantly, RTDs open the possibility for people who are BLV to be creators rather than just passive receivers of data visualisations. We want to make our own data visualisations from equations or data; or to import existing images and make sense of them. I have also enjoyed the experience of using a touch-sensitive RTD to draw by dragging my finger across the surface.

\subsection{Vision Specialist Teachers}
\emph{Timeliness:} As a teacher and a parent of a student with a vision impairment, I am most interested in RTDs enabling students who are BLV to access information at the same time as their peers. Currently, the delays in producing hard copy tactile graphics mean that students cannot access the curriculum at the same time as the rest of their classmates, limiting their participation in learning. Ideally, there would be a quick and easy way for classroom teachers to  transfer data and graphs into a meaningful representation on an RTD without needing expertise in braille or accessible formats.

\subsection{Accessible Formats Producers}
\emph{Immediate feedback and corrections:}
The successful creation of accessible graphics relies on translation and interpretation in order to simplify a diagram for touch while retaining the most important information. Since the advent of computerised accessible graphics production, there has been the added difficulty of designing for touch using a flat visual display. The introduction of RTDs could potentially make our task much easier, firstly because we are able to check our work tactually throughout the editing process and secondly because any corrections or refinements suggested by the user can be implemented immediately.

\emph{User creation:} Ultimately, we hope that RTDs could reduce our role dramatically by transferring power to the user, who should be able to generate and adjust their own tactile graphics, particularly in the realm of data visualisation.

\emph{Design guidelines:} I was at first sceptical about the utility of RTDs given the small number of pins available. However, the multiple pin heights available on a Graphiti RTD allowed a surprising amount of information to be conveyed. This required a new way of thinking about the design of tactile graphics, with new design guidelines required specifically for RTDs.

\subsection{Assistive Technology Providers}
\emph{User creation}: What most excites me about RTDs is the possibility that a blind person reviewing a tactile graphic may have an opportunity to edit it and create their own tactiles, perhaps with their finger position being tracked by sensors, or tracing shapes that pop up as the finger moves -- just as sighted people draw on iPad apps.

\emph{Dynamic content}: Being dynamic, the data on an RTD could change in real-time. For example, an RTD could be  connected to laboratory equipment to display measurement values as they change. On a boat, sailing speed, wind direction, and compass bearing could be displayed in real-time. 

\emph{Most likely use cases:} RTDs are heavy and expensive, so they may not be suitable for accessing maps or other graphics on the go. They are more suitable for high-use cases at a fixed location, for example at school or in the workplace. US funding models mean that educational use will likely be the first big market for these devices.

\section{Implications for the Visualisation Community}

Stakeholders are clear that RTDs offer a fundamentally new kind of tactile graphic experience, one that is the haptic equivalent of interactive data visualisation.
The effective introduction of RTDs must be informed by learnings from data visualisation, design guidelines for static tactile graphics, and co-design with people with lived experience of blindness. Key areas of inquiry include:

\begin{enumerate}
\item Designing tactile graphics with visualisation best practice in mind, trialling on RTDs and undertaking user testing. This will determine the optimal visualisations and \emph{use cases}. With a limited pins available, judicious design choices are essential and will be aided by new \emph{design guidelines} based on research into whether traditional visualisation design principles hold and where they need to be adapted.

\item The display of tactile graphics on RTDs is the first opportunity for BLV people to access \emph{dynamic content} and interact with tactile diagrams. The visualisation community can help to investigate effective interaction methods for features such as zooming and panning, as well as the combination of RTDs with speech and sonification for labelling.

\item Key understandings of visualisation creation should also be considered in the context of RTDs, in particular empowering BLV people as \emph{user creators}. For example, the first DotPad software does not allow for nonvisual graphics creation but DotInc have made their API available for developers to make their own apps for the creation, conversion and display of graphics on the DotPad. Other manufacturers have indicated that they would like to collaborate on software design and would benefit from the expertise of the visualisation development community. Such authoring systems will also be vital in exploiting the opportunities for \emph{timeliness}.
\end{enumerate}

The data visualisation community and their expertise will be crucial in supporting the accessibility community to explore the potential of RTDs and help ensure that people who are BLV can best benefit from the introduction of affordable RTDs in the near future.

\bibliographystyle{abbrv-doi}
\bibliography{RTDs}
\end{document}